\documentclass[compactaffiliation]{Interspeech}

\interspeechcameraready

\title{NGPU-LM: GPU-Accelerated N-Gram Language Model \\ for Context-Biasing in Greedy ASR Decoding}

\author[affiliation={1}]{Vladimir}{Bataev}
\author[affiliation={1}]{Andrei}{Andrusenko}
\author[affiliation={1}]{Lilit}{Grigoryan}
\author[affiliation={2}]{Aleksandr}{Laptev}
\author[affiliation={2}]{Vitaly}{Lavrukhin}
\author[affiliation={2}]{Boris}{Ginsburg}

\affiliation[]{}{NVIDIA}{Armenia}
\affiliation[]{}{NVIDIA}{USA}
\email{\{vbataev, aandrusenko, lgrigoryan, alaptev, vlavrukhin\}@nvidia.com}


\usepackage{hyperref}
\usepackage{comment}
\usepackage{algorithm}
\usepackage{algorithmic}
\usepackage{multirow}

\begin{document}

\maketitle

\begin{abstract}

Statistical n-gram language models are widely used for context-biasing tasks in Automatic Speech Recognition (ASR). However, existing implementations lack computational efficiency due to poor parallelization, making context-biasing less appealing for industrial use. This work rethinks data structures for statistical n-gram language models to enable fast and parallel operations for GPU-optimized inference. Our approach, named NGPU-LM, introduces customizable greedy decoding for all major ASR model types - including transducers, attention encoder-decoder models, and CTC - with less than 7\% computational overhead. The proposed approach can eliminate more than 50\% of the accuracy gap between greedy and beam search for out-of-domain scenarios while avoiding significant slowdown caused by beam search. 
The implementation of the proposed NGPU-LM is open-sourced.

\end{abstract}

\section{Introduction}

The emergence of end-to-end approaches has significantly simplified the development of Automatic Speech Recognition (ASR) systems. The first end-to-end ASR model, based on Connectionist Temporal Classification (CTC)~\cite{graves2006ctc}, allowed to learn the alignment between input acoustic features and target transcriptions without the time-consuming GMM-HMM training process used in conventional hybrid ASR systems~\cite{Hinton2012DeepNN}. Later, the introduction of the Recurrent Neural Transducer (RNN-T)~\cite{graves2012rnnt}, which incorporates an LSTM-based decoder module, enabled the model to better capture language dependencies. Advances in machine translation have further paved the way for adapting the Attention Encoder-Decoder (AED) approach~\cite{chan_las} to speech recognition, thereby improving recognition accuracy and integrating additional modalities (e.g., translation, time-stems prediction, etc.~\cite{Radford2022RobustSR,canary}) into a single model. 

Recently, Large Language Models have been applied to speech tasks (SpeechLM)~\cite{Zhang2023SpeechGPTEL,Rubenstein2023AudioPaLMAL}, showing promise as the next evolution in speech foundation models. However, training and deploying SpeechLM models remain highly resource-intensive, less stable, and still relatively underexplored.

End-to-end ASR models often exhibit a bias toward the training data domain, which leads to accuracy degradation when encountering out-of-domain test data. To mitigate this, context-biasing techniques are used to incorporate additional out-of-domain information through either deep or shallow fusion. Deep fusion involves feeding context data into the ASR model during training or fine-tuning ~\cite{Pundak2018DeepCE,Jain2020ContextualRF}, but this process can be computationally expensive. Similarly, SpeechLM supports context-biasing through prompting~\cite{Chen_2024}. However, this requires modifications to the training process (in-context learning) and does not scale well with an increasing number of target phrases. In contrast, shallow fusion-based methods~\cite{Zhao2019ShallowFusionEC,Jung2021SpellMN,Andrusenko2024FastCF} apply context biasing exclusively during the decoding stage, eliminating the need for costly model retraining.

The most common shallow fusion strategy is beam search decoding with an external language model (LM)~\cite{Zhao2019ShallowFusionEC}. While neural LMs based on Long Short-Term Memory (LSTM)~\cite{Hochreiter1997LongSM} or Transformer~\cite{Vaswani2017AttentionIA} architectures can serve as the external model, they require a substantial amount of out-of-domain text for training. In contrast, a statistical LM based on n-grams offers a simpler alternative, requiring significantly fewer resources and a smaller training dataset. One popular implementation of such a model is provided by the open-source KenLM library~\cite{heafield2011kenlm}, which represents the n-gram LM using a trie-based data structure optimized for CPU execution.

Although shallow fusion with an external LM improves out-of-domain recognition accuracy, beam search decoding introduces a significant slowdown compared to greedy decoding. This issue is particularly evident in Transducer-based ASR models, where beam search necessitates multiple evaluations of the Joint and Prediction network components. Moreover, the CPU-based implementation of the n-gram LM further restricts the decoding speed for other ASR architectures. Consequently, practitioners often face the need to choose between speed and accuracy in speech recognition.

Recent developments have introduced a more efficient greedy decoding method for Transducers based on the label-looping algorithm \cite{bataev2024labellooping} coupled with CUDA graphs~\cite{galvez24_speedoflight}. This advancement has significantly accelerated standard greedy decoding, widening the speed gap between the new GPU-based greedy approach and standard beam search.

In this paper, we present NGPU-LM, a novel GPU-accelerated implementation of a statistical n-gram language model. Our approach leverages a universal trie-based data structure that enables fast batched queries, thereby unlocking new possibilities for ASR decoding. NGPU-LM facilitates context-biasing in rapid greedy decoding on GPU across all major ASR architectures, including CTC, Transducers, and AED. The proposed method provides up to 10.6\% relative WER improvement while avoiding the substantial slowdown typically associated with beam search. 

The main contribution of our work is of two parts:
\begin{itemize}
    \item NGPU-LM: A novel data structure representing a statistical n-gram language model optimized for parallel computing;
    \item Shallow Fusion in Greedy Decoding: Enabling fast context-biased decoding across major ASR architectures (CTC, Transducers, AED) with extremely low overhead.
\end{itemize}
Our implementation is open-sourced in the NeMo~\cite{kuchaiev2019nemo} toolkit.

% \newpage
\section{Methods}

\subsection{Statistical LM Background}

\begin{figure}[t]
  \centering
  \includegraphics[width=8cm]{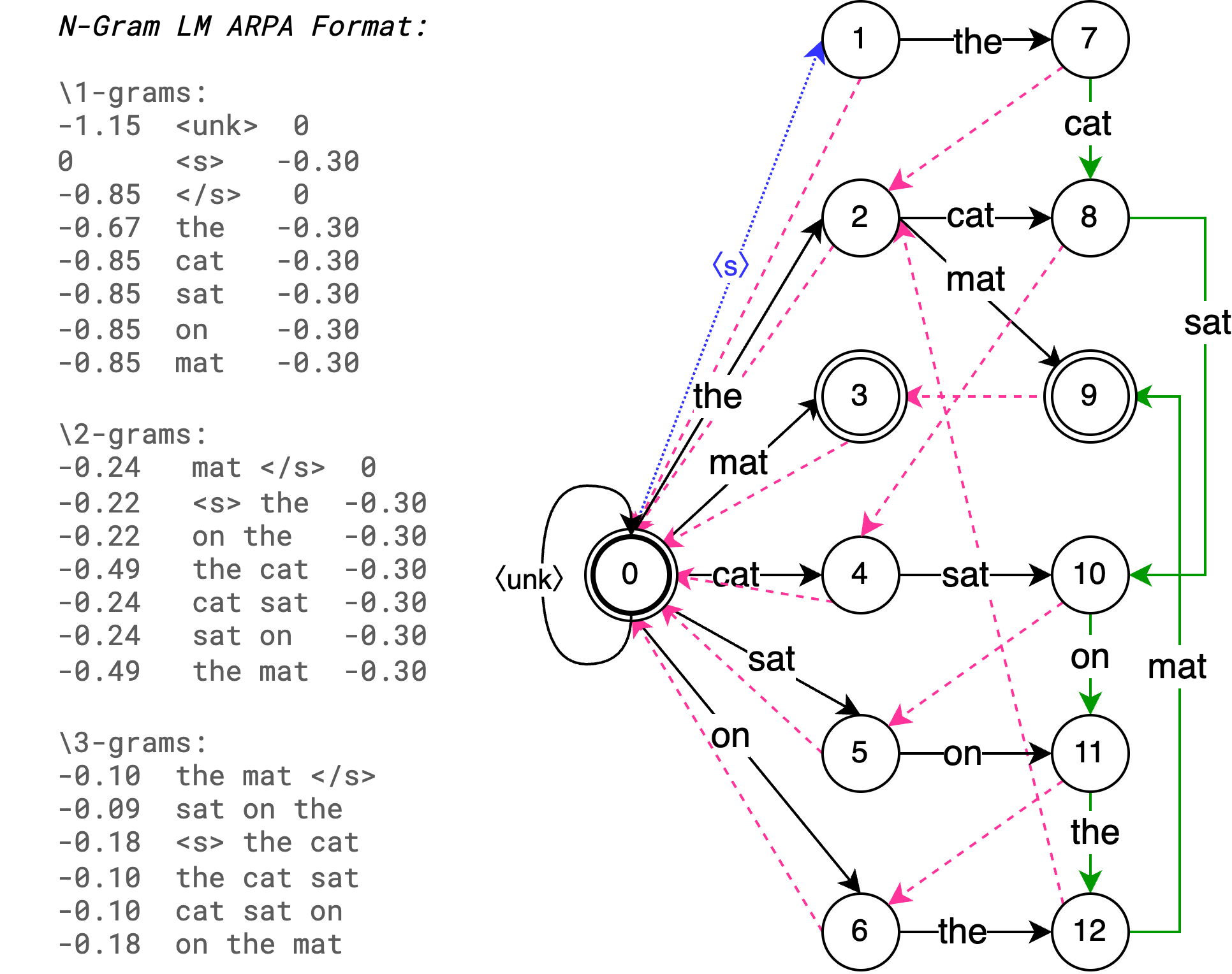}
  \caption{\textbf{3-gram LM} built from the text ``the cat sat on the mat''. 
    \textbf{Arrows}: \textbf{black} – unigram and bigram transitions; 
    \textbf{green} – 3-gram (highest order) transitions; \textbf{pink dashed} – backoff transitions.
    \textbf{Double circles}: final states.
    }
  \label{fig:ngram-tree}
  \vspace{-15pt}
\end{figure}

\begin{figure}[t]
\vspace{-15pt}
\begin{algorithm}[H]
   \newcommand\id[1]{\texttt{#1}}
   \caption{Inference of NGPU-LM}
    \label{algo:NGPU-LM}
\begin{algorithmic}[1]
    \footnotesize
    \REQUIRE $\id{state}$ - current LM state for hypothesis \\
    {\bfseries lm\_params (tensors):} \\
    \quad \textbf{arcs}: $\id{arc\_tokens}$, $\id{arc\_weights}$, $\id{arc\_to\_states}$ \\
    \quad \textbf{states to arcs ranges}: $\id{start\_arcs}$, $\id{end\_arcs}$ \\
    \quad \textbf{states to backoffs}: $\id{boff\_weights}$, $\id{boff\_to\_states}$ \\
    \STATE $\id{next\_states} = \text{full}\left([V], \text{fill\_value}=-1\right)$
    \STATE $\id{next\_scores} = \text{empty}\left([V]\right)$
    \STATE $\id{acc\_boff} = 0.0$ \hspace{6mm} \text{\# backoff weight accumulator}
    \WHILE{($\id{next\_states} == -1$).any()}
        \STATE $\id{scores} = \text{zeros}\left([V]\right)$
        \STATE $\id{to\_states} = \text{full}\left([V], \text{fill\_value}=-1\right)$
        \STATE \text{\# get arcs for the current state}
        \STATE $\id{s}, \id{e} = \id{start\_arcs}[\id{state}], \id{end\_arcs}[\id{state}]$
        \STATE $\id{tokens} = \id{arc\_tokens}[\id{s}:\id{e}]$
        \STATE $\id{scores[tokens]} = \id{acc\_boff} + \id{arc\_weights}[\id{s}:\id{e}]$
        \STATE $\id{to\_states}[\id{tokens}] = \id{arc\_to\_states}[\id{s}:\id{e}]$
        \STATE \text{\# fill in data for not yet observed tokens}
        \STATE $\id{mask} = \left(\id{next\_states} == -1\right)$  \text{\# not visited states mask}
        \STATE $\id{next\_scores}[\id{mask}] = \id{scores}[\id{mask}]$
        \STATE $\id{next\_states}[\id{mask}] = \id{to\_states}[\id{mask}]$
        \STATE \text{\# traverse by backoff transition }
        \STATE $\id{acc\_boff} \mathrel{+}= \id{boff\_weights}[\id{state}]$
        \STATE $\id{state} = \id{boff\_to\_states[\id{state}]}$
    \ENDWHILE
    \STATE {\bfseries return} $\id{next\_scores}$, $\id{next\_states}$
    \sffamily
\end{algorithmic}
\end{algorithm}
\vspace{-22pt}
\end{figure}

\textbf{N-gram LM Structure.} 
Statistical N-gram language model is a probabilistic model used to predict the likelihood of a sequence of text tokens (word- or subword-level). It is based on frequencies of subsequences (n-grams) observed in textual training data.
To explain the structure in depth, let's consider the word-level n-gram LM built from text and stored in a human-readable ARPA format, see Fig.~\ref{fig:ngram-tree}. Each line in an ARPA file describing the n-gram follows the format:
``$\mathrm{W(token|context)}$ $\mathrm{[context]}$ $\mathrm{token}$ $\mathrm{B(context+token)}$'',
where $\mathrm{context}$ is a sequence of $N-1$ tokens (context can be empty if $N=1$). Probabilities (weights $W$) are stored as logarithms, allowing all calculations to be performed using summation instead of multiplication. Backoff weights $B$ are used for discounting probabilities when transitioning to a smaller context. In addition to regular tokens, n-gram models include special symbols such as start-of-sequence $\langle s \rangle$, end-of-sequence $\langle /s \rangle$, and the unknown token unigram $\langle unk \rangle$. 

To score the sentence, the algorithm starts with the start-of-sequence token as a context and, iteratively, for each token, accumulates the log-probabilities for $\mathrm{W(token | context)}$. If the model does not contain $\mathrm{W(token | context)}$ explicitly, the algorithm iteratively reduces the context, going through backoffs until $\mathrm{W(token | context_{reduced})}$ is found, following the formula $\mathrm{W(token | context) = B(context) + W(token | context_{1:})}$. If the start state is reached (empty context), and $token$ is not found, then $\mathrm{\langle unk \rangle}$ weight is assigned.

\textbf{Graph Representation.} 
In practice, n-gram LM is represented as a graph, specifically a suffix tree. Prefixes (forward state chains) represent n-grams. \textit{States} represent context. \textit{Suffix links} represent fallbacks to a shorter context (backoff transitions).
Fig.~\ref{fig:ngram-tree} visualizes such representation for 3-gram word LM built from the sentence \textit{``the cat sat on the mat''}. Note that for 3-gram transitions, the arcs point directly to 2-gram states since no separate states are needed for the highest-order n-grams.
To simplify storage, models often omit explicit representation of the $\langle s \rangle$ and $\langle /s \rangle$ tokens. Instead, the start-of-sequence is treated as a separate state, while the end-of-sequence $\langle /s \rangle$ acts as a final weight marker (double-border circle in the figure).

Standard graph implementations, such as CPU-focused libraries like KenLM~\cite{heafield2011kenlm} and parallel computing toolkits like gLM~\cite{Bogoychev2016NGramLMParallel}, are designed to efficiently handle queries of the form ``$\mathrm{(token, state) \rightarrow (token\_weight, next\_state)}$''.
To achieve high performance, queries $\mathrm{P(token|context)}$ are typically processed in one of two ways~\cite{heafield2011kenlm}, using sorted transition arrays to enable binary search (or K-ary for parallel implementations), or by employing hash tables for $O(1)$ transition lookups.

However, when probabilities need to be aggregated across the entire vocabulary (we use BPE~\cite{bpe} tokens in our experiments), performing separate queries for each token is computationally expensive. Instead, we can determine the range of transitions to a given state and fill in the probabilities for tokens present in that state in parallel. Secondly, since the length of backoff chains is limited by the order of the LM (up to 10 in our experiments), the number of iterations required to aggregate probabilities over the full vocabulary remains relatively small. These two insights enable the construction of a simple yet highly efficient data structure and parallelized algorithm for full-vocabulary batched queries.

\subsection{NGPU-LM Data Structure.} 
For efficient greedy decoding with full vocabulary scoring, we need to support queries like 
$\mathrm{state[B] \rightarrow (token\_weights[B, V], next\_states[B, V])}$,
where the input is a batch of states (representing a batch of hypotheses), and the output is the probability distribution produced by LM for the whole vocabulary $|V|$ for each of state in the batch, along with the corresponding next states.
To achieve this, we store all the data in tensors, specifically:
\begin{itemize}
    \item \textit{Arcs (transitions) data tensors}: tokens, weights, source states ($\mathrm{from\_states}$), target states ($\mathrm{to\_states}$)
    \item \textit{State data tensors}: backoff target states ($\mathrm{boff\_to\_states}$), backoff weights, and final weights.
\end{itemize}

For the start (unigram) state, we store transitions for all vocabulary tokens to avoid extra checks during queries. If the token is absent in ARPA, we use normalized $\mathrm{\langle unk \rangle}$ weight, eliminating the need for a separate unknown token representation. This ensures that retrieving probabilities for zero-context queries (unigrams) is straightforward.

To optimize retrieval, we sort all arcs by the tuple ``$\mathrm{(from\_state, token)}$'', and introduce two additional tensors, $\mathrm{start\_arcs}$ and $\mathrm{end\_arcs}$, which define the range of arcs for each state. This allows for efficient weight retrieval using a simple parallelized ``index-select'' operation. Specifically, for non-batched queries for state $s$, we first get $\mathrm{(start, end)}$ transitions boundaries and then assign scores and next states.
To obtain scores over the full vocabulary, the algorithm iteratively follows backoff transitions, aggregating the weights for present tokens until all scores are found. The maximum number of iterations does not exceed the order of the model.
The full non-batched algorithm is listed in Algorithm~\ref{algo:NGPU-LM}. It can be easily extended to support the input of the batch of states. We are omitting the batched version here due to space limits and encourage the readers to check the complete implementation\footnote{\begin{scriptsize}\url{https://github.com/NVIDIA/NeMo/pull/10989}\end{scriptsize}}.

\textbf{Implementation and Acceleration.}
The baseline algorithm is implemented in PyTorch. However, loop conditions are evaluated on CPU (due to eager mode constraints), slowing down execution. This fact also makes the implementation incompatible with CUDA graphs, limiting decoding speed optimizations. To overcome these issues, we implemented a custom kernel in Triton~\cite{tillet2019triton}, which retains the algorithm's clarity while significantly improving speed. Our PyTorch implementation remains available as a fallback option for environments where Triton cannot be used (e.g., CPU-based decoding).

\subsection{Greedy Decoding with NGPU-LM}

In typical implementations of the beam search decoding~\cite{watanabe2018espnet,kuchaiev2019nemo} with LM, to extend the hypothesis, the algorithm first selects top-k token expansions for the hypothesis and then the LM scores these tokens, retrieving respecting weights one by one. 
% We refer to this approach as ``early pruning'' since LM is used to re-weight only already pruned hypotheses. 
This approach is not applicable for greedy decoding since, on each step, only one expansion is selected, and LM is not useful. To enable the use of LM in greedy mode, we need to use full vocabulary scoring, first adding language model weight to the expanded hypotheses in each step and then applying greedy selection. Such approach would be impractically slow with conventional LM implementations due to CPU-bound sequential queries but is efficient with NGPU-LM.

\textbf{Transducers.} 
We integrate NGPU-LM into the decoding pipeline for conventional transducer (RNN-T)~\cite{graves2012rnnt} and the Token-and-Duration Transducer (TDT)~\cite{xu2023tdt}. Our implementation is built on the greedy label-looping~\cite{bataev2024labellooping} algorithm accelerated with CUDA graphs~\cite{galvez24_speedoflight} to ensure maximum decoding efficiency. 
In initial experiments, we found that directly combining transducer and language model scores led to an increased deletion rate. This issue arises due to the dominance of the blank symbol (which is not scored by LM), negatively impacting performance. To address this, we introduce a two-stage token selection algorithm that maintains a distinction between blank and non-blank predictions at each decoding step. Firstly, we do standard greedy token prediction. If the $\langle blank \rangle$ symbol is predicted, we retain it. Otherwise, we rescore all non-blank probabilities using the language model and apply greedy selection among non-blank symbols. This approach allows for an increase in language model weight effectively, leading to significant performance improvements, especially for out-of-domain.

\textbf{CTC.} 
For CTC model decoding, we adopt a similar approach to transducer-based decoding but introduce an additional distinction between symbol categories at each step. Specifically, we preserve three groups: ``blank'', ``repeated (previous) token'' and ``other tokens''. Since repeated tokens are collapsed due to the CTC decoding rule, they should not be rescored by the language model. To accelerate sequential autoregressive processing with LM, which involves tiny operations with significant kernel launch overhead, we apply CUDA graphs similar to the approach used for transducers~\cite{galvez24_speedoflight}.

\textbf{Attention Encoder-Decoder.} 
Unlike transducer and CTC models, encoder-decoder models with attention do not use the ``blank'' symbol. Instead, they introduce an end-of-sentence ``$\langle eos \rangle$'' symbol, which directly corresponds to the final weight in the language model. The final weight is either directly listed in n-grams or is accessible by backoff transitions. 
To optimize final weight retrieval during decoding, we precompute and store final weights for all states at the model loading time by traversing backoff transitions.
The preprocessing step significantly simplifies and accelerates the decoding. We integrate NGPU-LM into both greedy and beam search decoding.

\section{Experimental Setup}

\subsection{ASR modeling}

We evaluated four ASR architectures: CTC, Transducers (RNN-T and TDT), and AED. For all models, we employed the FastConformer encoder \cite{gulati2020conformer,rekesh2023fastconformer}, configured with 108M parameters. The RNN-T and TDT models utilize a single-layer LSTM decoder with 640 hidden units, resulting in 114M parameters in total, while the AED model features a 4-layer Transformer decoder with a hidden size of 1024 (179M total model params).

To simulate varying resource conditions, we experimented with two training datasets: a large corpus of approximately 24k hours of English speech from various open sources and the 960-hour Librispeech dataset \cite{Panayotov2015LibrispeechAA}, representing a low-resource scenario. Text data was tokenized using a BPE tokenizer~\cite{bpe} with a vocabulary of 1024 tokens. All models were trained under identical conditions using the NeMo~\cite{kuchaiev2019nemo} framework.

To assess model performance, we tested on out-of-domain datasets: SLURP (command-focused)~\cite{bastianelli2020slurp} and SPGISpeech (finance-related)~\cite{ONeill2021SPGISpeech} (SPGI). We built\footnote{\begin{scriptsize}\href{https://github.com/NVIDIA/NeMo/blob/v2.3.1/scripts/asr_language_modeling/ngram_lm/train_kenlm.py}{\texttt{github.com/NVIDIA/NeMo/.../train\_kenlm.py}}\end{scriptsize}} token-level 10-gram language models (without pruning, except for SPGI\footnote{\begin{scriptsize}
\texttt{ngram\_prune=[0,0,0,0,1,1,2,2,3,3]}
\end{scriptsize}}) using KenLM toolkit~\cite{heafield2011kenlm} for these domains using their respective training data. The development sets were then used to fine-tune the fusion weights for the external language models. All models require less than 100MB on the GPU memory, with the exception of the model built on SPGI, which uses 1.5GB.
For the ASR model trained on LibriSpeech data, we additionally evaluate its performance on the WSJ \texttt{eval92} dataset~\cite{paul1992wsj}, using \texttt{dev93} to tune the language model weight.
For beam search with transducers, we use Modified Adaptive Expansion Search~\cite{maes} with KenLM~\cite{heafield2011kenlm} model fusion. We use Flashlight~\cite{kahn2022flashlight} decoder for CTC models. In all beam search runs, $beam=4$ is used.

\textbf{Factorized and stateless Transducers.} 
There are a number of studies that have explored factorized transducers \cite{McDermott2019ADR,Meng2020InternalLM,mhat}, where the primary idea is to decouple label probabilities from the blank token. This separation facilitates the estimation of the internal language model (ILM) and its subtraction during the shallow fusion with an external LM. In our work, we investigate the integration of the proposed NGPU-LM for Hybrid Autoregressive Transducer (HAT) \cite{hat}, using the same training dataset of 24k hours.
Additionally, we examine a stateless decoder \cite{stateless} that replaces the conventional LSTM-based decoder with embeddings of the last two predicted tokens. The lightweight decoder may be less biased toward the training data, making it more suitable for shallow fusion with an external LM.

\textbf{Metrics.}
We measure decoding accuracy using the Word Error Rate (WER) metric and assess decoding speed with the inverse Real-Time Factor (RTFx). RTFx is calculated as the total audio duration divided by the evaluation time of the ASR model. RTFx is measured after a single warm-up run on a single A6000 GPU with a batch size of 32.

\section{Results}

\begin{table}[t]
    \centering
    \caption{
    Decoding Fast Conformer Large models (108M Encoder parameters) trained on large data (24k hours) on out-of-domain data (SLURP, SPGI). WER vs RTFx.
    \textit{RTFx} is the inversed real-time factor (higher is better).
    \vspace{-8pt}
    }
    \resizebox{\columnwidth}{!}{
    \begin{tabular}{l c | c c | c c }
    \toprule
    \multirow{2}{*}{\textbf{Decoding}} & \multirow{2}{*}{\textbf{LM}} & \multicolumn{2}{c|}{\textbf{SLURP}} & \multicolumn{2}{c}{\textbf{SPGI}} \\
    % \cmidrule(l){3-6}
    & & \textbf{WER} \%$\downarrow$ & \textbf{RTFx}$\uparrow$ & \textbf{WER} \%$\downarrow$ & \textbf{RTFx}$\uparrow$  \\
    \midrule
    \textbf{RNN-T} \\
    \midrule
    greedy & - & 22.47 & 996 & 5.93 & 2352   \\
    greedy & NGPU-LM  &  20.36 & 973 & 5.43 & 2230   \\
    beam & - & 22.34 & 87 & 6.26 & 95 \\
    beam & LM & 19.34 & 79 & 4.73 & 87  \\
    \midrule
    \textbf{TDT} \\
    \midrule
    greedy &  & 23.48 & 1076 & 6.45 & 2877   \\
    greedy & NGPU-LM  & 21.21 & 1063 & 5.91 & 2798  \\
    beam & - & 22.99 & 95 & 6.29 & 101 \\
    beam & LM  & 19.48	& 83 & 4.89 & 88  \\
    \midrule
    \textbf{AED} \\
    \midrule
    greedy &  & 25.06 & 732 & 6.10 & 785   \\
    greedy & NGPU-LM  & 23.59 & 722 & 5.48 & 736  \\
    beam & - & 24.31 & 244 & 6.08 & 292 \\
    beam & NGPU-LM & 19.68 & 442 & 4.65 & 286 \\
    \midrule
    \textbf{CTC} \\
    \midrule
    greedy &   & 24.88 & 1175 & 6.75 & 3311   \\
    greedy & NGPU-LM  & 22.66 & 1152 & 6.34 & 3108  \\
    beam & - & 24.88 & 941 & 6.75 & 1905 \\
    beam & LM & 20.32 & 838 & 4.42 & 1294  \\
    \bottomrule
    \end{tabular}
    }
    \label{tab:decoding-high-resource}
    \vspace{-8pt}
\end{table}

\begin{table}[t]
    \centering
    \caption{
    Decoding Fast Conformer Large models (108M Encoder parameters) trained on LibriSpeech dataset (960h). WER~\%. Test data: SLURP, SPGI, WSJ. 
    }
    \vspace{-8pt}
    \resizebox{\columnwidth}{!}{
    \begin{tabular}{l c c | c c c}
    \toprule
    \textbf{Model} & \textbf{Decoding} & \textbf{LM} & \textbf{SLURP} & \textbf{SPGI} & \textbf{WSJ}  \\
    \midrule
    \multirow{3}{*}{RNN-T} & greedy & - & 48.66 & 14.65 & 7.25  \\
    & greedy & NGPU-LM  & 44.90 & 13.10 & 6.36 \\
    & beam & LM & 44.53 & 12.88 & 6.75 \\
    \midrule
    \multirow{3}{*}{CTC} & greedy & - & 51.06	& 16.09 & 7.71  \\
    & greedy & NGPU-LM  &  47.58 & 14.41 & 6.47  \\
    & beam & LM & 42.67 & 11.51 & 5.14 \\
    \bottomrule
    \end{tabular}
    }
    \label{tab:decoding-low-resource}
    \vspace{-15pt}
\end{table}

\begin{table}[t]
    \centering
    \caption{
    Comparison of RNN-T and HAT models with LSTM and Stateless decoders depending on language model setup.
    }
    \vspace{-8pt}
    \resizebox{\columnwidth}{!}{ % TODO: remove?
    \begin{tabular}{c | c | c | c c c}
    \toprule
    \multirow{2}{*}{\textbf{Model}} & \multirow{2}{*}{\textbf{Decoder}} & \textbf{LM} & \multicolumn{3}{c}{\textbf{WER \%}} \\
    \cmidrule(l){4-6}
     &  & \textbf{Setup} & \textbf{SLURP} & \textbf{SPGI} & \textbf{Internal} \\
    \midrule
    \multirow{4}{*}{RNN-T} & \multirow{2}{*}{LSTM} & - & 22.47 & 5.93 & 11.81 \\
     & & +LM & 20.36 & 5.43 & \textbf{9.84} \\
    \cmidrule(l){2-6}
     & \multirow{2}{*}{Stateless} & - & 22.90 & 5.86 & 11.90 \\
     & & +LM & 20.91 & \textbf{5.41} & 10.23 \\
    \midrule
    \multirow{6}{*}{HAT} & \multirow{3}{*}{LSTM} & - & 22.96 & 5.99 & 12.41 \\
     & & +LM & 20.82 & 5.42 & 10.17 \\
     & & -ILM+LM & \textbf{19.79} & 5.47 & 10.12 \\
    \cmidrule(l){2-6}
     & \multirow{3}{*}{Stateless} & - & 23.14 & 5.97 & 12.45 \\
     & & +LM & 21.20 & 5.43 & 10.46 \\
     & & -ILM+LM & 20.42 & 5.50 & 10.19 \\
    \bottomrule
    \end{tabular}
    }
    \label{tab:decoding-rnnt-ablation}
    \vspace{-15pt}
\end{table}

Table~\ref{tab:decoding-high-resource} presents the decoding performance of models trained on sufficiently large datasets. The results demonstrate that integrating NGPU-LM into greedy decoding significantly improves accuracy while incurring less than 7\% computational overhead across all models.

\textbf{RNN-T and TDT Models.}
With the proposed greedy decoding using NGPU-LM, we achieve relative word error rate reduction (WERR) ranging from 8.4\% (both models on SPGI) to 9.7\% (TDT on SLURP). Notably, beam search without an LM fails to provide a significant accuracy improvement while causing a substantial 11x performance degradation. This indicates that the LM itself is not a bottleneck in beam search. On SLURP, our greedy decoding approach accounts for over 56\% of the relative improvement provided by beam search and more than 30\% on SPGI.

\textbf{AED.}
For the AED model, we observe WERR improvements between 5.9\% (SLURP) and 10.2\% (SPGI) when using NGPU-LM in greedy mode. Interestingly, when applying beam search without an LM on SLURP, we notice abnormal performance degradation due to the model's inability to generate the end-of-sequence symbol. This issue is resolved by LM fusion.

\textbf{CTC models.}
CTC models also benefit from NGPU-LM, achieving 6.1\% to 8.9\% WERR improvements. However, the performance gap between the proposed greedy decoding and beam search is the largest among all model types.

\textbf{Low-Resource Scenario.}
Table~\ref{tab:decoding-low-resource} presents results for models trained on smaller data, showing trends similar to high-resource cases. In this setting, the proposed approach recovers more than 87\% of the gap between greedy and beam search for the RNN-T model with up to 10.6\% WERR improvement on SLURP and SPGI, and outperforms beam search on WSJ data. For the CTC model, it covers approximately 35\% of the gap.

\textbf{Factorized and stateless Transducers.}
The comparison results of RNN-T and HAT models with different decoders are presented in Table \ref{tab:decoding-rnnt-ablation}. HAT\textsubscript{LSTM} showed the best result on SLURP, but slightly inferior to RNN-T\textsubscript{LSTM} in the case of SPGI. The use of a stateless decoder did not show a clear advantage of this approach in the conditions of LM fusion.
In addition, we evaluated our internal dataset designed to test keywords recognition in the computer science domain. The language model for this set was built on the list of target keywords consisting of 200 elements. The obtained results showed the effectiveness of the proposed method in the conditions of keyword boosting with the best result for the RNN-T\textsubscript{LSTM} model.

\section{Conclusion}

We introduced a novel highly parallelized representation of n-gram language models, enabling fast, batched, full-vocabulary queries. This approach seamlessly integrates into any ASR decoding pipeline. Additionally, we introduce efficient language model fusion for greedy decoding for all types of ASR models with minimal computational overhead. Our results demonstrate that this customization significantly improves out-of-domain performance demonstrating up to 10.6\% WERR improvement while avoiding the high computational cost of beam search, making it particularly valuable for industry-scale ASR systems. In future work, we plan to explore deeper integration of our model into various beam search algorithms, further enhancing decoding efficiency and accuracy.

\ifinterspeechfinal
% \section{Acknowledgements}
% Acknowledgement should only be included in the camera-ready version, not in the version submitted for review. 
\else
     
\fi

% \newpage
% \clearpage 
\bibliographystyle{IEEEtran}
\bibliography{mybib}

\end{document}